\documentclass[preprint,aps]{revtex4}

\usepackage{graphicx} 
\usepackage{dcolumn}
\usepackage{bm}
\usepackage{color}
\usepackage{amsmath} 
\usepackage{natbib}

\begin{document}
\bibliographystyle{apsrev}

\title{On the relevance of uncorrelated Lorentzian pulses for the interpretation of turbulence in the edge of magnetically confined toroidal plasmas} 
\preprint{Submitted to Physical Review Letters}

\author{B.Ph.~van Milligen$^1$, R.~S\'anchez$^2$, and C.~Hidalgo$^1$}

\affiliation{$^1$Laboratorio Nacional de Fusi\'on, Asociaci\'on EURATOM-CIEMAT, 28040 Madrid, SPAIN\\
$^2$Universidad Carlos III de Madrid,  Legan\'es, 28911 Madrid, SPAIN}
\date{\today}

\begin{abstract}
Recently, it has been proposed that the turbulent fluctuations measured in a linear plasma device could be described as a superposition of  uncorrelated Lorentzian pulses with a narrow distribution of durations, which would provide an explanation for the reported quasi-exponential power spectra.
Here, we study the applicability of this proposal to edge fluctuations in toroidal magnetic confinement fusion plasmas. 
For the purpose of this analysis, we  introduce a novel wavelet-based pulse detection technique that offers important advantages over existing techniques. 
It allows extracting the properties of individual pulses from the experimental time series, and quantifying the distribution of pulse duration and energy, as well as temporal correlations. 

We apply the wavelet technique to edge turbulent fluctuation data from the W7-AS stellarator and the JET tokamak, and find that the 
pulses detected in the data do not have a narrow distribution of durations and are not uncorrelated.
Instead, the distributions are of the power law type, exhibiting temporal correlations over scales much longer than the typical pulse duration.
These results {suggest that turbulence in open and closed field line systems may be distinct and} cast doubt on the proposed ubiquity of exponential power spectra in this context.
\end{abstract}

\maketitle

{
The understanding of plasma turbulence is of vital importance for the development of fusion as an alternative energy source, in view of the fact that it is the dominant mechanism behind the unwanted energy loss of confined plasmas, referred to as power degradation.}
Over the last three decades, the {complex} dynamics of turbulent transport in magnetically confined toroidal 
{devices such as tokamaks} and stellarators have
been probed experimentally by examining {edge} fluctuation data measured with Langmuir probes. 
{It was found that edge fluctuations} exhibit self-similar properties in certain regimes, leading to power spectra characterized by power law sections extending over a meaningful range of frequencies~\cite{Carreras:1998b,Carreras:1999d,Rhodes:1999,Xu:2004,Xu:2005}. 

Recently, Maggs et al.~\cite{Maggs:2011} have proposed that this type of results should be reinterpreted in terms of an uncorrelated superposition of Lorentzian pulses with a narrow distribution of pulse durations. 
The resulting power spectrum would be predominantly exponential, 
which would fit many of the spectra reported in literature~\cite{Pace:2008,Hornung:2011,Pedrosa:1999b}. 
According to the authors, a consequence of this proposal is that the underlying dynamics would be low-dimensional quasi-deterministic chaos.
{This view appears to deny the importance of the mutual interaction between confinement and turbulence, widely considered a vital ingredient in the understanding of power degradation in fusion plasmas, and contrasts} with the hypothesis of self-organized critical dynamics  proposed in the past~\cite{Dendy:1997,Carreras:1997}.

In this Letter, we {study} this proposal using edge fluctuation data from {a tokamak and a stellarator}.
As the pulse distribution is central to the issue at hand,
we introduce a novel wavelet decomposition technique that allows recovering its complete statistical properties. 
{The analysis clarifies} that there are significant differences between {turbulence in linear plasmas} and {toroidal} fusion plasmas, at least in the operational regime examined (L-mode).

We {hold} that the power spectrum of a signal {is not} determined by the spectrum of the individual pulse, unless it is composed of uncorrelated pulses with identical duration. Any experimental signal $s(t)$ can be approximated by a sum of pulses: 
\begin{equation}\label{Lorentzsum}
s'(t) = \sum_{i=1}^N{\frac{a_i}{\sqrt{\tau_i}} L \left (\frac{t-t_i}{\tau_i} \right )}.
\end{equation}
Here, $L(t)$ is the basic pulse shape, and $t_i$, $i=1,\dots,N$ is a set of times at which pulses of amplitude $a_i$ and duration $\tau_i$ occur.
For discretely sampled data, the absolute difference between the signal $s$ and the sum $s'$ can be made arbitrarily small, provided $L(t)$ satisfies some minimal requirements.
For fixed $\tau_i = \overline{\tau}$, Eq.~\eqref{Lorentzsum} can be rewritten as
\begin{equation}\label{convolution}
s'(t) = \int_{-\infty}^\infty{\left [ \sum_{i=1}^N{\frac{a_i \delta(t'-t_i)}{\sqrt{\overline{\tau}}}}\right ]L\left (\frac{t-t'}{\overline{\tau}}\right ) dt'},
\end{equation}
such that $s'(t)$ is a convolution of the pulse shape $L(t)$ and a function (in square brackets) specifying the amplitude and time of the pulses. Consequently, the power spectrum of $s'$ is the product of spectrum of this function and the spectrum of the pulse shape.

It is useful to choose $L(t)$ as similar as possible to the typical shape of actual observed pulses in the data series -- 
which can in principle be determined using techniques such as conditional averaging~\cite{Boedo:2001,Boedo:2003,Xu:2005,Teliban:2007}.
Following Maggs et al.~\cite{Maggs:2011}, one could use $L(t) = 1/\pi(1+t^2)$ (Lorentzian).
However, an alternative pulse shape such as $G(t) = 1/(1+e^{|t|})$ provides an equally good fit to the experimental pulse shape obtained from our data using conditional averaging.
While the spectrum of $L(t)$ is exponential ($\hat L(\omega) \propto e^{-|\omega|}$), the spectrum of $G$ is power-law ($\lim_{\omega \to \infty} \hat G(\omega) \propto  \omega^{-2}$).
This seems to suggest that the pulse shape does not predetermine the shape of the experimental power spectrum.

{\bf Wavelet analysis.} Next, we introduce a wavelet technique to obtain the statistics of and the correlation between pulses of an arbitrary time signal, with potential application in wide variety of fields. We construct a (`mother') wavelet by combining two pulses with opposite sign:
$\psi_1(t) = L(t) - \frac12L\left (\frac{t}{2}\right)$.
By analogy with the well-known Difference Of Gaussians (DOG) wavelet, we call this new wavelet the Difference Of Lorentzians (DOL) wavelet. Its Fourier transform is $\hat \psi_1(\omega) = (\exp(-|\omega|)-\exp(-2|\omega|))/\sqrt{2\pi}$ (essentially, exponential decay).
For this wavelet, the admissibility constant~\cite{Daubechies:1992}
$C_\psi = \frac12 \int_{-\infty}^{\infty}{{|\hat \psi_1(\omega)|^2}/{|\omega|}~d\omega}$
satisfies $0 < C_\psi < \infty$, leading to a proper wavelet transform.
This makes the full power of continuous wavelet analysis available to the issue at hand~\cite{Milligen:1999}.
We define the wavelet by
$\psi_\tau(t) = \frac{1}{\sqrt{\tau}} \psi_1 \left ( \frac{t}{\tau} \right )$,
such that its $L^2$-norm does not depend on $\tau$.
The wavelet transform and its inverse are
\begin{align}
a(t,\tau) &= \int{s(t') \psi_\tau(t-t')dt'}, \label{forward} \\
s'(t) &= \frac{1}{C_\psi} \int\!\!\!\int{a(t',\tau)\psi_\tau^*(t-t') \frac{d\tau dt'}{\tau^2}}.  \label{inverse}
\end{align}
This allows decomposing any (square integrable) signal $s(t)$ into sums of Lorentzians with arbitrary precision~\cite{Daubechies:1992} -- in particular, any such signal $s(t)$ with any given spectrum.
Consequently, a signal composed of a sum of Lorentzians, Eq.~\eqref{Lorentzsum}, {\it can have an arbitrary spectrum}. 

The conditions to obtain an exponential power spectrum from the pulse distribution are easily gleaned from Eq.~\eqref{inverse}. 
By setting all wavelet coefficients $a$ to zero except those corresponding to a specific scale $\overline \tau$
(formally, $a(t,\tau) = a(t,\overline \tau)\delta(\tau-\overline \tau)$), 
the expression for the reconstructed signal $s'$ becomes a convolution, and 
{the power spectrum $P_{s'} = |\hat s'|^2$ can be written:}
\begin{equation}
P_{s'}(\omega) = P_a(\omega)\cdot P_\psi(\omega),
\end{equation}
{Thus}, provided the spectrum of $a(t,\overline \tau)$ is essentially flat (i.e., $a$ is random, uncorrelated),
the spectrum of $s'$ will have the same shape as the wavelet spectrum (i.e., essentially exponential). 
Note that {a similar} conclusion can be obtained directly from Eq.~\eqref{convolution}{, without invoking wavelets}.
{Thus,} it is not sufficient to require only that the signal consists of Lorentzian pulses with a fixed duration in order to produce an exponential spectrum.
This {is} easily demonstrated by generating a signal based on the following simple algorithm:
(1) Choose a number $N$ of pulse amplitudes $a_i$, $i=1,...,N$. 
(2) Choose $N-1$ time intervals $\Delta t_i$.
(3) Construct the pulse time array from $t_i = t_{i-1} + \Delta t_{i-1}$.
(4) Compute the signal $s(t)$ from Eq.~\eqref{Lorentzsum}, using $\tau_i=\overline \tau$. The resulting signal is a sum of fixed-duration Lorentzian pulses.

\begin{figure}
  \includegraphics[trim=0 0 0 0,clip=,width=12cm]{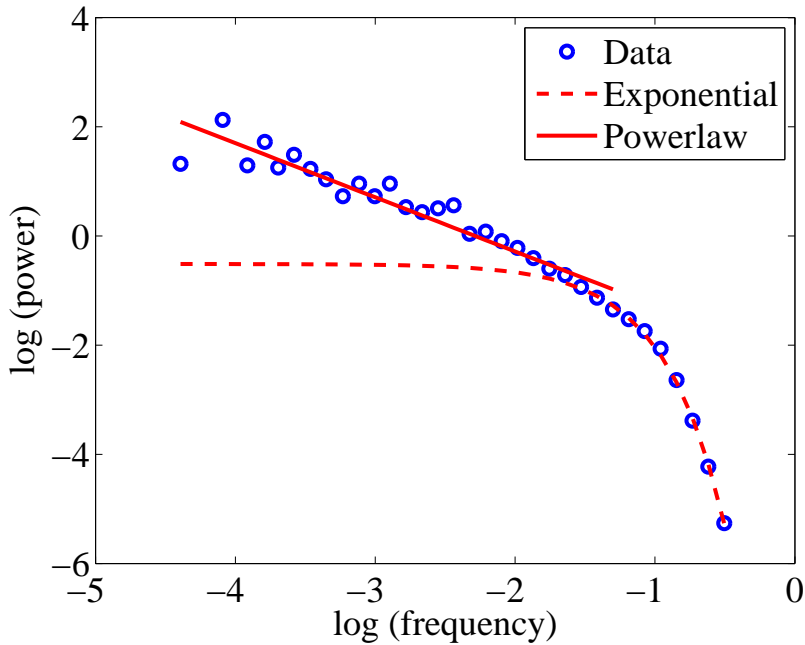}
\caption{\label{pulsetrain}Spectrum for artificial data according to Eq.~\eqref{Lorentzsum}, constructed using the algorithm described in the text, showing that even a signal built from fixed-duration Lorentzian pulses is not always completely exponential.}
\end{figure}

By way of example, we set $N=5000$ and $\overline \tau = 3$. 
For $\Delta t_i$ we take a uniformly distributed random variable in the range $[0,10)$, and for $a_i$ a fractional Gaussian noise (fGn) with a certain degree of persistence (Hurst parameter $H=0.8$)~\cite{Mandelbrot:1969}. 
{The spectrum of this artificial signal is} shown in Fig.~\ref{pulsetrain}. {It clearly follows a power-law} for low frequencies (extending over nearly 3 orders of magnitude of $f$), {while at high frequencies it falls of exponentially} (as expected).
While we do not pretend this simulation to be a model of any physical system, it certainly shows that a signal consisting of a sum of fixed duration Lorentzian pulses does not possess an exponential spectrum {\it unless the pulses are uncorrelated in time.}

{\bf Analysis of stellarator/tokamak turbulent fluctuation data.}
The wavelet transform, Eq.~\eqref{forward}, offers an efficient technique for identifying pulses in experimental data $s(t)$.
For this purpose, the transform $a(t,\tau)$ is computed and the local maxima of $|a(t,\tau)|$ are found~\cite{Misiti:2007}.
The set of values $a_{\rm max}, t_{\rm max}, \tau_{\rm max}$ indicate, respectively, the amplitude, time of occurrence, and duration of pulses in the data series. {This} technique is capable of identifying all individual pulses, regardless of their scale or amplitude, unlike, e.g., conditional or threshold techniques -- assuming, of course, that the wavelet shape $\psi_1$ adequately matches the experimental pulse shape and that pulses do not overlap.

\begin{figure}
  \includegraphics[trim=0 0 0 0,clip=,width=12cm]{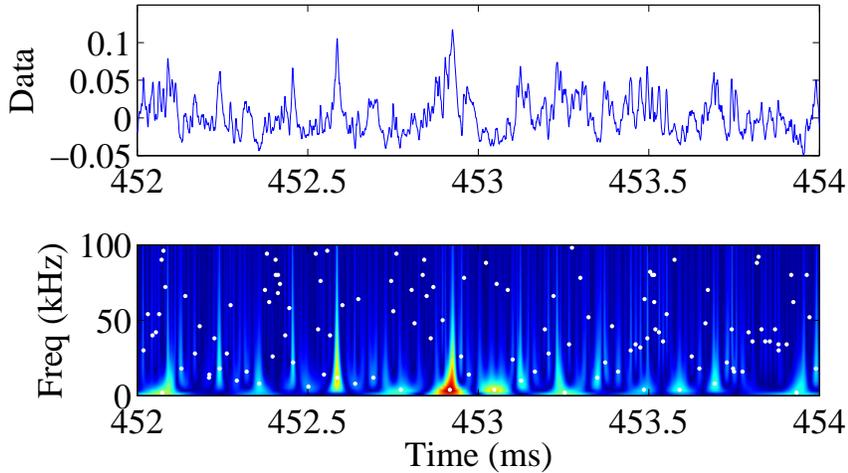}
\caption{\label{W7_wavelet}
Top: a short time section (2 ms) of $I_{\rm sat}$ data.
Bottom: wavelet transform $|a(t,\tau)|$ (blue: small amplitude; red: large amplitude). The vertical axis shows $f=1/\tau$; only the lowest frequencies are shown ($f < 100$ kHz), although the spectrum is computed up to the Nyquist frequency.
White spots indicate the detected pulses (local maxima of $|a(t,\tau)|$).
}
\end{figure}

To demonstrate the technique and test the hypothesis of Maggs et al, Fig.~\ref{W7_wavelet} shows a wavelet spectrum for Langmuir probe data (the ion saturation current, $I_{\rm sat}$) obtained in the plasma edge of W7-AS, discharge 35425, $430 \le t \le 455$ ms~\cite{Carreras:1998b}.
These data are obtained just inside the last closed magnetic surface in an L-mode discharge.

\begin{figure}
  \includegraphics[trim=0 0 0 0,clip=,width=12cm]{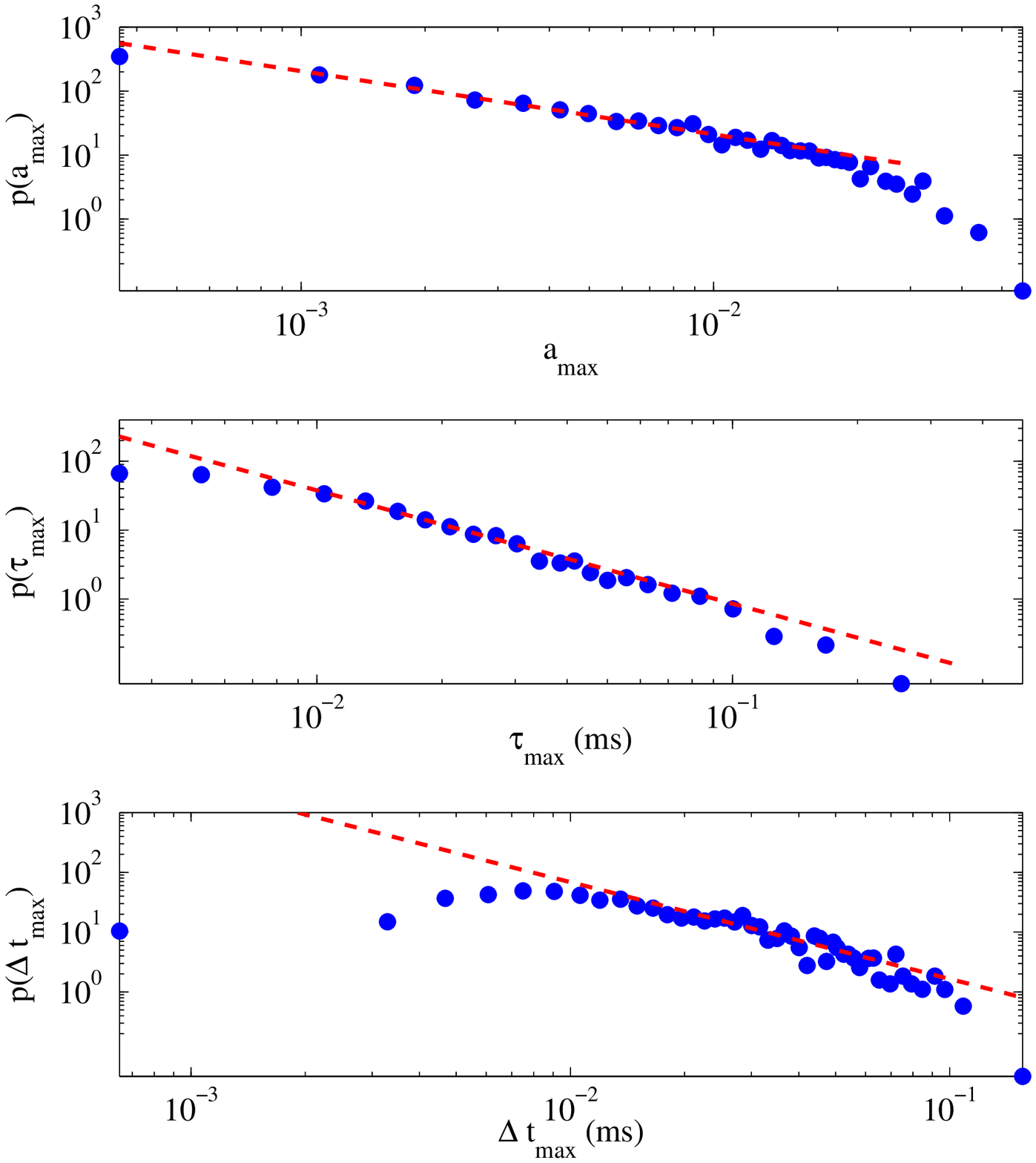}
\caption{\label{W7_distrib}
Analysis of the pulse distribution obtained from wavelet analysis (W7-AS).
Top:  pulse amplitude $p(a_{\rm max})$. Center: pulse duration $p(\tau_{\rm max})$. Bottom: waiting time between energetic pulses $p(\Delta t_{\rm max})$.
}
\end{figure}

Fig.~\ref{W7_distrib} shows the probability distributions of the (absolute) amplitude $p(a_{\rm max})$, the pulse duration $p(\tau_{\rm max})$, and the waiting time between pulses $p(\Delta t_{\rm max})$. 
To calculate the waiting time distribution, only energetic pulses are considered (i.e., pulses satisfying $E = (a_{\rm max}/\tau_{\rm max})^2 > E_\theta$, where $E_\theta$ is an appropriate threshold~\cite{Farge:2006}).
All show a power law structure. 
It is clear that neither the assumption of a narrow pulse duration distribution, nor that of an uncorrelated pulse train appear justified
(the latter would imply a Poisson {waiting time distribution}).

\begin{figure}
  \includegraphics[trim=0 0 0 0,clip=,width=12cm]{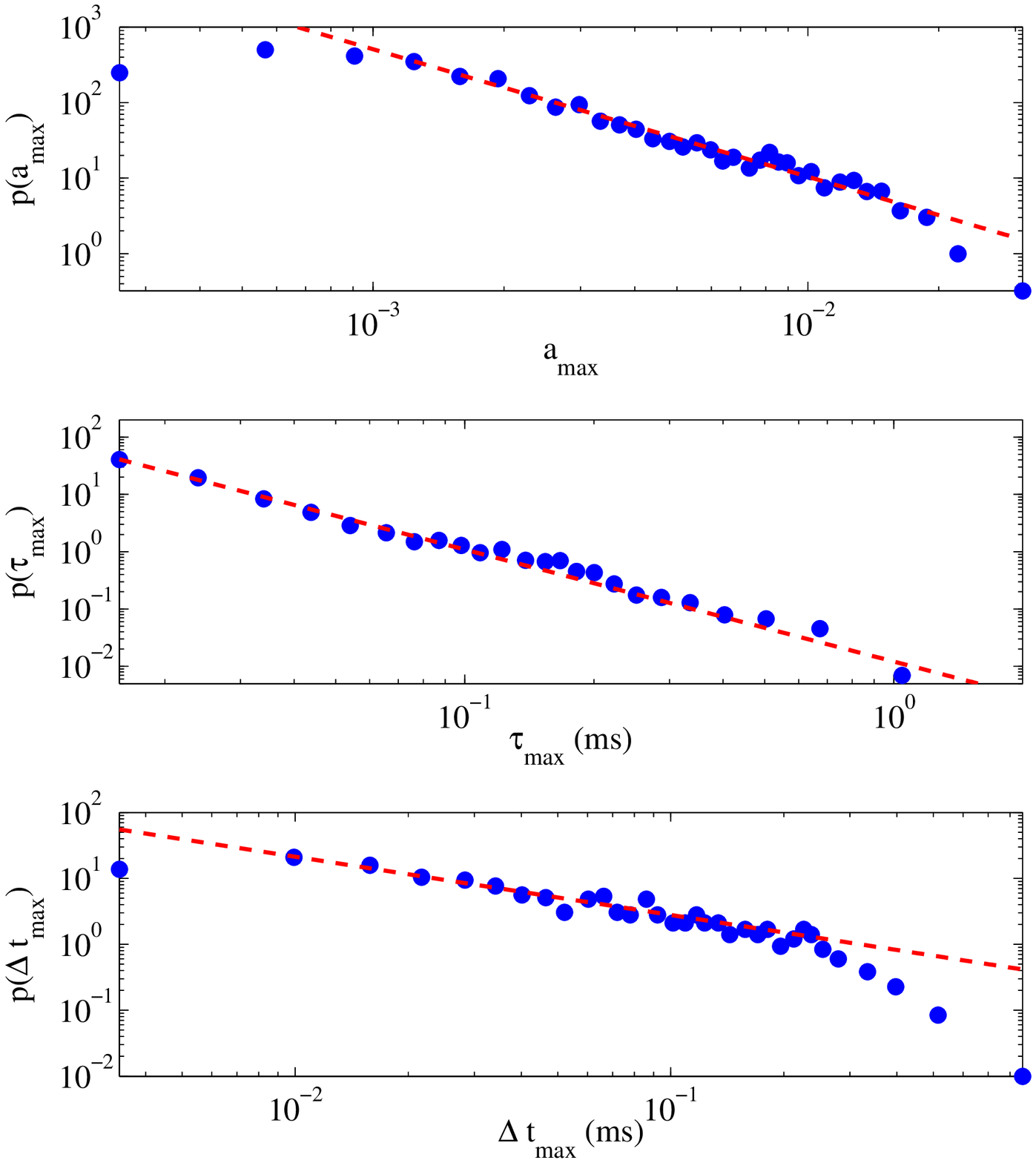}
\caption{\label{JET_distrib}
Analysis of the pulse distribution obtained from wavelet analysis (JET).
Top:  pulse amplitude $p(a_{\rm max})$. Center: pulse duration $p(\tau_{\rm max})$. Bottom: waiting time between energetic pulses $p(\Delta t_{\rm max})$.
}
\end{figure}

Fig.~\ref{JET_distrib} shows similar results for {$I_{\rm sat}$} measurements in the edge of a JET plasma (discharge 54008, $62.52 \le t \le 62.58$ s)~\cite{Hidalgo:2005}.
We draw attention to the significant length of the observed power laws.

{\bf Discussion.}
In previous work, the shape of experimental spectra in the edge of fusion devices was studied, 
and power laws were detected over a range of frequencies in certain conditions 
(typically, measurements obtained in L-mode discharges at positions either inside or outside of, but very close to, the last closed surface)
~\cite{Carreras:1998b,Carreras:1999d,Rhodes:1999,Xu:2004,Xu:2005}. 
{Many} factors may affect the shape of the spectra, such as the Doppler effect due to plasma rotation, the presence of modes, and the position of the probe in the plasma, making its study rather non-trivial.

The analysis presented here, based on a novel technique that is more powerful than techniques used in the past,
provides deeper insight into the {complex} nature of the dynamics in toroidal magnetic confinement fusion plasmas:
the distribution of pulse amplitudes and durations follows a power law, and long-range temporal correlations (memory effects) are significant.
Memory effects refer to persistence over times much longer than the turbulence autocorrelation time,
presumably caused by the self-organization of profiles and turbulence~\cite{Sanchez:2005}.
The existence of such memory effects is also supported convincingly by a number of studies involving techniques such as the rescaled range (Hurst parameter) and cross correlation~\cite{Carreras:1998b,Carreras:1999b,Carreras:1999c,Gilmore:2002,Carralero:2011}, the structure function~\cite{Yu:2003,Budaev:2008}, {and} quiet time analysis~\cite{Sanchez:2002b,Sanchez:2003}.
In these studies, temporal correlations {are} reported for an ample set of toroidal fusion devices~\cite{Carreras:1998b}, 
{and} it was concluded that memory effects can exceed the local decorrelation time by up to three orders of magnitude.

{\bf Conclusions.}
In this work, we have examined the claim put forward in~\cite{Maggs:2011} that pulse shapes and spectra of fluctuations in the edge of fusion plasmas might be explained by a sequence of Lorentzian pulses, associated with exponential {spectra.}
To address this issue, a new pulse detection technique was developed
{based on a} wavelet closely matching the typical observed pulse shapes.
{This technique} may find application in {other fields} where complex dynamics are thought to play an important role.

The analysis has revealed that {exponential spectra} are only obtained from Lorentzian pulses {if}:
(i) the {distribution of pulse durations is narrow} and 
(ii) the temporal distribution of {pulses} has a flat spectrum (implying no temporal correlations).
{The distribution of pulse amplitudes and durations calculated for actual fusion plasma edge data is broad while temporal correlations are strong, in} accordance with earlier work~\cite{Carreras:1998b,Carreras:1999b,Carreras:1999c,Gilmore:2002,Yu:2003,Budaev:2008,Sanchez:2002,Sanchez:2003}, and invalidating both these {requirements}.

Thus, we conclude that the dynamics of turbulence in the edge of magnetically confined toroidal plasmas are often quite different from and richer than in linear plasmas. 
{In linear devices, it seems likely that parallel losses along the field lines to end plates are so fast that they limit} the range of times over which temporal correlations can be established, whereas in toroidal plasmas the existence of closed magnetic surfaces { allows phenomena with longer time scales to manifest themselves.}
{In this situation, the} interaction between profiles and turbulence (plasma self-organization) may {then} be a source of long-range temporal correlations~\cite{Sanchez:2005}, which may require conditions of good confinement and strong gradients.

Although the results reported here seem to be typical for tokamaks and stellarators, it cannot be excluded that there are specific regimes and/or regions in toroidal plasmas that might exhibit dynamics more similar to the results reported by Maggs et al at LAPD-U. For instance, the decorrelation between pulses should be faster in a region of strong poloidal sheared flows or in the far scrape-off layer, where magnetic field lines are connected to the first wall.

{\bf Acknowledgements:}
Research sponsored in part by DGICYT of Spain under project Nrs. ENE2009-07247 and ENE2009-12213-C03-03.


\end{document}